\documentclass[
showkeys,
reprint,
superscriptaddress,
nofootinbib,
nobibnotes,
longbibliography,
aps,
prb,
floatfix,
nofootinbib
]{revtex4-1}

\usepackage{mathptmx}
\usepackage{empheq}

\usepackage{amsmath}
\usepackage{amsfonts}
\usepackage{amssymb}
\usepackage{dsfont}
\usepackage{bm}
\usepackage[hyperfootnotes=false,breaklinks=true,colorlinks=true,linkcolor=blue,urlcolor=blue,citecolor=blue]{hyperref}
\usepackage[dvipsnames]{xcolor}

\newcommand{\E}{\mathcal E}
\newcommand{\F}{\mathcal F}
\newcommand{\Lie}{\mathcal L}

\newcommand{\LCM}{\nabla}

\newcommand{\SFS}{\chi}

\newcounter{example}[section]
\newcounter{remark}[section]
\newcounter{theorem}[section]
\newcounter{proposition}[section]
\newcounter{lemma}[section]
\newcounter{corollary}[section]
\newcounter{definition}[section]

\def\theremark{\arabic{section}.\arabic{remark}}
\def\thetheorem{\arabic{section}.\arabic{theorem}}

\def\thedefinition{\arabic{section}.\arabic{definition}}

\renewcommand*{\email}[1]{\footnote{Electronic address: \href{mailto:#1}{\nolinkurl{#1}} }}

\begin{document}

\title{Constructing massive particles surfaces in static spacetimes}
\author{Igor Bogush\email{igbogush@gmail.com}}
\affiliation{Moldova State University, strada Alexei Mateevici 60, 2009, Chi\c{s}in\u{a}u, Moldova}
\author{Kirill  Kobialko\email{kobyalkokv@yandex.ru}}
\author{Dmitri Gal'tsov\email{galtsov@phys.msu.ru}}
\affiliation{Faculty of Physics, Moscow State University, 119899, Moscow, Russia}

\begin{abstract}
The procedure for constructing the massive particle surfaces in static space-times is described in detail and the equivalence of the main results with the results of the geodesic approach is demonstrated.  
\end{abstract}


\maketitle

\setcounter{page}{2}

\setcounter{equation}{0}
\setcounter{subsection}{0}

\section{Introduction}
The increased interest in the theoretical description of black hole shadows\cite{Perlick:2004tq,Perlick:2021aok,Bronzwaer:2021lzo,Cunha:2018acu} has stimulated the development of new geometric tools applicable in stationary axisymmetric spacetimes. The standard approach appeals to the analysis of geodesic curves in the vicinity of black holes near regions where closed photon orbits are located, for example, unstable circular closed orbits of radius $r=3M$ in the Schwarzschild metric or spherical orbits in the Kerr field.\cite{Virbhadra:1999nm,Teo:2020sey} In both cases, such orbits fill certain hypersurfaces in spacetime, which can be described in geometric terms of submanifold theory.\cite{Chen} Namely, the circular orbits of photons in the first case form a photon sphere,\cite{Virbhadra:1999nm,Virbhadra:2002ju} which is an umbilic hypersurface, the first and second fundamental forms of which are proportional.\cite{Claudel:2000yi} Spherical photon surfaces in the Kerr metric fill \textit{partially umbilic surfaces},\cite{Kobialko:2020vqf} for which the proportionality of two fundamental forms is valid only on the subspace of the tangent space corresponding to a fixed value of the impact parameter. Together, these hypersurfaces fill a four-dimensional volume of spacetime known as the \textit{photon region}.\cite{Kobialko:2020vqf,grenzebach2016shadow,Grenzebach:2015oea,Grenzebach:2014fha} In many cases, studying these submanifolds in terms of surface geometry is simpler than studying the full geodesic structure, leading to a useful alternative description of black hole shadows.\cite{Kobialko:2023qzo}

In the presence of plasma,\cite{Perlick:2015vta,Perlick:2017fio,Briozzo:2022mgg,Perlick:2023znh,Bezdekova:2022gib} the orbits of photons are no longer lightlike geodesics, but rather geodesics followed by massive particles. The theory of corresponding \textit{massive particles surfaces} (MPS) was proposed in Ref. \cite{Kobialko:2022uzj} (see also Ref. \cite{Song:2022fdg}). Their description turned out to be possible also in terms of partially umbilical surfaces, depending, in addition, on a conserved quantity associated with the timelike Killing vector. The concept of the massive particle surface was reformulated in Ref.,\cite{Bogush:2023ojz} where the condition of fixed specific energy was replaced by a quadratic condition for two linear integrals of motion. In this case, Kerr spacetime contains massive particles surface, where massive particles can have different energies and azimuthal angular momenta subject to a quadratic condition. The massive particles surfaces can be extended taking into account electric and magnetic charges.\cite{Kobialko:2022uzj,Bogush:2023ojz} Note, that in the case of variable mass, the application of formalism \cite{Kobialko:2022uzj,Bogush:2023ojz} requires carrying out the Weyl transformation as discussed in Ref.\cite{Kobialko:2023qzo} Certainly, the entire theory can also be directly generalized to this case without this trick.

It is important, that the usual geodesic description of strong gravitational lensing and shadows leads to results identical to the description in terms of characteristic surfaces, but the second method not only provides simplifications and a deeper geometric picture, but leads to a variety of new theoretical tools for the analysis of uniqueness theorems,\cite{Yazadjiev:2015hda,Rogatko:2016mho,Cederbaum:2015fra,Cederbaum:2014gva,Rogatko:2024nzq} hidden symmetries,\cite{Kobialko:2021aqg,Kobialko:2022ozq} Penrose inequalities,\cite{Shiromizu:2017ego,Feng:2019zzn,Yang:2019zcn} etc. In particular, the connection between surfaces and hidden symmetries allows us to obtain in a closed form the contour of the shadow cast by a black hole.\cite{Kobialko:2023qzo} 

This method has recently been tested on several examples under both general relativity and modified gravity, and some discrepancies with the results of standard geodesic methods have been reported in Ref.\cite{Junior:2024cts}. We do not agree with this and show the complete equivalence of both approaches in all the examples discussed in the article. To avoid future misunderstandings, we consider it useful to discuss in more detail the application of our method in static spacetime.

The paper plan is the following. In Sec. \ref{sec:mps} we briefly describe the general construction method of massive particle surfaces introduced in Ref.\cite{Kobialko:2022uzj}. In Sec. \ref{sec:static} we present a detailed application of the general formalism to static space and derive general analytical expressions for the energy of massive particle surfaces. In Sec. \ref{sec:examples} we apply the developed formalism to many important particular examples and compare the results with the geodesic approach. The appendix explains some conventions adopted in the article.

\section{Massive particles surface (MPS)}\label{sec:mps}
Consider a stationary $n$-dimensional spacetime with metric $g_{\alpha\beta}$ possessing a timelike Killing vector $k^\alpha$  and the electromagnetic field described by a covector field $A_\alpha$ sharing the same symmetry,  $\Lie_{k} A_\alpha=0$. Let $\gamma$ denote an arbitrary particle worldline with an associated $n$-velocity $\dot{\gamma}^\alpha$ and obeys the following equations  
\begin{equation}
   \dot{\gamma}^\alpha \LCM_{\alpha}\dot{\gamma}^\beta =q F^\beta{}_{\lambda}\dot{\gamma}^\lambda , \quad \dot{\gamma}^\alpha\dot{\gamma}_\alpha=-m^2.  \label{eq_particles}
\end{equation} 
We can define the total $\E$, kinetic $\E_k$ and potential $\E_p$ energies of the particle as follows
\begin{equation}
    \E_k = -k_\alpha\dot{\gamma}^\alpha,\qquad
    \E_p = -qk_\alpha A^\alpha,\qquad
    \E = \E_k + \E_p.
\end{equation}
The electromagnetic field tensor reads $F_{\alpha\beta} = \partial_\alpha A_\beta - \partial_\beta A_\alpha$. This convection is self-consistent, but may differ in the sign of charge from some other references, see Appendix A. 

By definition,\cite{Kobialko:2022uzj} an MPS is an $n-1$-dimensional timelike hypersurface $S$ with a spacelike outer normal $n^\alpha$ such that any particle with a fixed energy $\E$, starting with an initial velocity lying in $S $ remains in $S$ forever.
This property can be expressed in terms of the induced metric tensor and extrinsic curvature tensor of $S$ defined as\cite{Kobialko:2022uzj}
\begin{equation}
    h_{\alpha\beta} = g_{\alpha\beta} - n_\alpha n_\beta,\qquad
    \chi_{\alpha\beta} = h_\alpha^\mu h_\beta^\nu \nabla_\mu n_\nu.
\end{equation}
Denote projection of the timelike Killing vector onto $S$ as $\kappa^\alpha = h_{\beta}^{\alpha}k^\beta$. Here we will consider only the case when the Killing vector is tangent to $S$, i.e., $\kappa^\alpha = k^\alpha$. In this case, the MPS is static.

The thorough derivation of conditions satisfied by an MPS for particles of mass $m$, charge $q$ and  fixed conserved energy $\E$, can be found in Ref.  \cite{Kobialko:2022uzj}. A crucial role is played by the following relation between the induced metric  the extrinsic curvature and the electromagnetic field tensor:
\begin{equation}\label{eq:condition_3}
\SFS_{\alpha\beta}=\cfrac{\SFS_\tau}{n-2} H_{\alpha\beta}+(q/\E_k)\F_{\alpha\beta},
\end{equation}
where
\begin{align}\label{eq:definitions_3} 
    &
    H_{\alpha\beta}=h_{\alpha\beta}+ (m^2/\E_k^2)  \kappa_\alpha \kappa_\beta,\qquad
    \F_{\alpha\beta}=\frac{1}{2}n^\rho F_{\rho (\alpha} \kappa_{\beta)}.
\end{align}
Note that for applications to particles of variable mass, such as photons in plasma, a Weyl transformation must be performed to make the mass constant before using expressions (\ref{eq:condition_3}) and (\ref{eq:definitions_3}).\cite{Kobialko:2023qzo} For massless and uncharged particles the relation (\ref{eq:condition_3}) reduces to the usual umbilicity condition, and the proportionality coefficient $\chi_\tau$ has the meaning of the trace of the second fundamental form accurate to the integer multiplier. For non-zero $m$ and/or $q$ only partial umbilicity holds \cite{Kobialko:2020vqf} in the following sense. Denote the system of linearly independent tangent vectors of $S$ orthogonal to $\kappa^\alpha$ as $\tau_{(i)}^\alpha, i=1,...,n-2$. One can derive from (\ref{eq:condition_3}) and (\ref{eq:definitions_3}) a  partial umbilicity and block-diagonality conditions in the form
\begin{subequations}
\begin{align} \label{eq:partially_umbilical}
    &
 \tau_{(i)}^\alpha \tau_{(j)}^\beta \SFS_{\alpha\beta} = 
    \frac{\SFS_\tau}{n-2}\tau^\alpha_{(i)}\tau^\beta_{(j)}h_{\alpha\beta},
    \\ &\label{eq:blok}
    n^\alpha \tau_{(i)}^\beta \left(
        \kappa^{-2}  \LCM_\beta\kappa_\alpha + \frac{1}{2}(q/\E_k) F_{\alpha \beta}
    \right) = 0.
\end{align}
\end{subequations}
The first of these equations means that the proportionality of the first and second fundamental forms is valid only for their values on tangent vectors orthogonal to the Killing vector. Then, the proportionality coefficient $\chi_\tau$ depends on the $\tau$-directions only. This value is important for the entire calculation. 

Equations (\ref{eq:condition_3}) and (\ref{eq:definitions_3}) allows us to determine the MPS for a given energy $\E$ without using the geodesic equations. However, instead, it is more convenient to determine the corresponding energy for a given umbilical surface. The reason is that the equations for energy turn out to be much simpler and are expressed by a quadratic polynomial.
This can be done using  the  master equation derived in Ref. \cite{Kobialko:2022uzj} (Eq. (35)) which also follows from Eqs. (\ref{eq:condition_3}) and (\ref{eq:definitions_3}):
\begin{align} \label{eq:master}
    - \kappa^{-2} \kappa^\alpha n^\beta \nabla_\alpha \kappa_\beta =
    \frac{1 + (m/\E_k)^2\kappa^2}{n-2}\SFS_\tau +q\F/\E_k.
\end{align}
Resolving this equation with respect to energy, in the case of the tangent timelike Killing vector, we get (see Eq. (36) in Ref. \cite{Kobialko:2022uzj}):
\begin{equation}\label{eq:energy}
    \E_\pm =
    \pm m \sqrt{
              \frac{ \kappa^2 \SFS_\tau}{K}
            + \frac{\mathcal{F}^2 (n-2)^2 q^2}{4m^2K^2}
        }
    + \frac{\mathcal{F} (n-2) q}{2 K}
    - q\kappa_\alpha A^\alpha,
\end{equation}
where $\kappa^2$ is the squared norm of the $S$-projection of the Killing vector and
\begin{subequations}
\label{eq:K}
\begin{align}
    &
    K = - \SFS_\tau + \frac{n-2}{2} n^\alpha \nabla_\alpha \ln \kappa^2,
    \\ &
    \F = {\F^\alpha}_\alpha = n^\rho F_{\rho\lambda} \kappa^\lambda.
\end{align}
\end{subequations}
It is important that the right-hand side of Eq. (\ref{eq:energy}) must be constant on the surface, otherwise the surface under consideration is not an MPS. In order to obtain an expression for the "radius" of a massive particles surface, for a given energy $\E$ it is necessary to resolve Eq. (\ref{eq:energy}) as an implicit one. Note that the condition $d\E/dr = 0$ with respect to an appropriately defined radial variable $r$ distinguishes the marginally stable orbits, separating stable and unstable orbits. This condition was obtained under a number of additional assumptions (see Ref. \cite{Kobialko:2022uzj}), therefore, its use must be careful. However,  these assumptions are satisfied in all examples we consider below.

The limit $m=q=0$ corresponding to the photon surface leads to divergent specific energy $\E/m$ in Eq. (\ref{eq:energy}) since we will get $K=0$. Taking a step back, Eq. (\ref{eq:master}) degenerates to another condition on the surface $$\SFS_\tau = -(n-2) \kappa^{-2} \kappa^\alpha n^\beta \nabla_\alpha \kappa_\beta,$$ which does not depend on energy since null geodesics are conformally invariant. This is  identical to condition $K=0$.

\section{Axially symmetric static four-dimensional spacetimes} \label{sec:static}

Consider a general static axially symmetric four-dimensional spacetime with the following metric tensor
\begin{equation}\label{eq:metric_ansatz}
    ds^2 = -\alpha dt^2 + \lambda dr^2 + \beta d\theta^2 + \gamma d\phi^2,
\end{equation}
where $\alpha, \beta, \gamma, \lambda$ are functions of $r,\theta$ and the chosen timelike Killing vector is $k^\alpha\partial_\alpha=\partial_t$ (i.e., $k^2 = -\alpha$) and the condition $\Lie_{\partial_t} A_\alpha=0$ is imposed on the vector potential. We choose the ansatz for the massive particle surface in the form $r = \text{const}$. The induced metric $h_{\alpha\beta}$ corresponds to (\ref{eq:metric_ansatz}) with $dr=0$, i.e. 
\begin{equation}\label{eq:metric_S}
ds^2|_S = -\alpha dt^2 + \beta d\theta^2 + \gamma d\phi^2. 
\end{equation}
The extrinsic curvature tensor reads \cite{Kobialko:2022uzj}
\begin{equation} \label{eq:chi_ansatz}
    \chi_{\alpha\beta} dx^\alpha dx^\beta = \frac{1}{2\sqrt{\lambda}}\left(
        -\partial_r \alpha dt^2 + \partial_r \gamma d\phi^2 + \partial_r \beta d\theta^2
    \right).
\end{equation}
In this case, the Killing vector is tangent to the surface $\kappa^\alpha = k^\alpha$ and two other tangent directions orthogonal to the Killing vector are $\tau_{(\theta)}^\alpha=\delta^\alpha_\theta$, $\tau_{(\phi)}^\alpha=\delta^\alpha_\phi$. Using Eqs. (\ref{eq:metric_S}), (\ref{eq:chi_ansatz}), the condition (\ref{eq:partially_umbilical}) reads
\begin{equation}
  \frac{ \partial_r \gamma}{\sqrt{\lambda}} = 
    \SFS_\tau\gamma ,
    \qquad    
    \frac{\partial_r \beta}{\sqrt{\lambda}} = 
   \SFS_\tau\beta,
\end{equation}
resulting in
\begin{equation}\label{eq:con_SFS}
   \SFS_\tau = \frac{1}{\sqrt{\lambda}} \partial_r \ln \beta, 
   \qquad
   \partial_r (\beta / \gamma) = 0,
\end{equation}
and the second condition (\ref{eq:blok}) leads to $F_{r\phi}=F_{r\theta}=0$. Accordingly, one can calculate the following quantities:
\begin{subequations}
\begin{equation}  \label{eq:K_ex}
    K =- \frac{1}{\sqrt{\lambda}}  \partial_r \ln (\beta/\alpha),
\end{equation}
\begin{equation} 
\F = n^\rho F_{\rho\lambda} \kappa^\lambda=\frac{1}{\sqrt{\lambda}} F_{rt},
\end{equation}
\begin{equation}\label{eq:energy_g}
    \E_\pm =
    \pm m \sqrt{
              \frac{ \alpha  \cdot \partial_r  \ln \beta}{  \partial_r \ln \beta/\alpha}
            + \frac{ F^2_{rt} }{\left(\partial_r \ln \beta/\alpha\right)^2} \frac{q^2}{m^2}
        }
    - \frac{F_{rt} }{ \partial_r \ln \beta/\alpha} q
    - q A_t.
\end{equation}
\end{subequations}
All conditions collected together are
\begin{equation}
\partial_r (\beta/\gamma) = 0, \quad F_{r\phi}=F_{r\theta}=0, \quad \Lie_{\partial_t} A_\alpha=0.
\end{equation}
In particular, in the absence of the electric charge $q=0$, conditions on the electromagnetic field can be omitted, and from (\ref{eq:energy_g}) follows
\begin{equation}  \label{eq:energy_q_zero} 
  \E_\pm^2/m^2 =
    \frac{ \alpha  \cdot \partial_r  \ln \beta}{  \partial_r \ln \beta/\alpha}, \quad \partial_r (\beta/\gamma) = 0.
\end{equation}  
Furthermore, for the photon sphere $q=m=0$, the condition $K=0$ leads to the well known total umbilicity: $\partial_r \ln \alpha=\partial_r \ln \beta=\partial_r \ln \gamma$.

In comparison with Ref. \cite{Junior:2024cts} we have gotten significantly different expressions (\ref{eq:con_SFS}) and (\ref{eq:energy_q_zero}) (compare with Eqs. (57) and (112) in Ref. \cite{Junior:2024cts}; we use notations $\alpha=A$, $\lambda=B$, $\beta = C$, $\gamma = C \sin^2\theta$ and signature $(-,+,+,+)$). Most likely, the discrepancy between these results appeared due to the erroneous calculation of the average curvature $\SFS_\tau$ in the umbilical sector in Ref. \cite{Junior:2024cts} which is given in the article without justification by the following expression (Eq. (57) in Ref. \cite{Junior:2024cts})
\begin{equation}\label{eq:tau_wrong}
\SFS_\tau \overset{\text{?}}{=} 2\sqrt{\alpha/\beta},
\end{equation} 
which clearly differs from (\ref{eq:con_SFS}). Expression (\ref{eq:tau_wrong}) coincides with the correct answer if one assumes the relation $2\sqrt{\alpha\beta\lambda}=\partial_r \beta$, which takes place in the Schwarzschild-like ansatz $\alpha=\lambda^{-1}$ and $\beta=r^2$. The Reissner-Nordström metric falls under this special case,  but not the conformally transformed metric and Culetu model considered in Ref. \cite{Junior:2024cts}. Keeping this in mind, there is no surprise in discrepancy obtained in Ref. \cite{Junior:2024cts} between the geodesic approach and surface approach based on erroneous expression (\ref{eq:tau_wrong}). Further, we will show that making use of the correct expression (\ref{eq:con_SFS}) leads to results completely consistent  with the geodesic approach. 

\section{Examples} \label{sec:examples}

Examples considered in Ref. \cite{Junior:2024cts} and also Refs.  \cite{Pugliese:2013xfa,Heydari-Fard:2021pjc} include Reissner-Nordström metric in electrovacuum model, electrically charged dilatonic black holes, Schwarzschild metric in conformal gravity, effective metric in Culetu model, which is general relativity coupled to non-linear electrodynamics. These models cover different directions of modified gravity stirring up interest to their analysis within the MPS approach. 

\subsection{Schwarzschild-like ansatz}
The generic Schwarzschid-like ansatz of spherically symmetric metrics can be written in the form
\begin{equation}
    \alpha=\lambda^{-1}=f(r),\qquad
    \beta=r^2,\qquad
    \gamma=r^2\sin^2\theta,
\end{equation}
Applying Eq. (\ref{eq:energy_q_zero}), the energy takes a simple form
\begin{equation} \label{eq:sch_e}
  \E_\pm^2/m^2 =
    \frac{2 f^2}{2 f-r f'}.
\end{equation}  
In order to determine the innermost stable circular orbits, we will use the condition of marginally stable orbits $d\E/ dr = 0$. We will also assume the following three natural conditions. First, $\E \neq 0$ because particles are timelike or null. Second, $f$ has a finite non-zero value, otherwise such points would correspond to event horizons or singularities. Third, the ISCO radius does not coincide with the radius of the photon surface. Using these three conditions allows us to simplify the condition on ISCO orbits  as follows:
\begin{equation}\label{eq:sch_de}
    3 f f' + r f f'' - 2 r f'^2
    = 0.
\end{equation}
Since $f$ is considered to possess a finite non-zero value, the equation on the photon surface follows from the denominator setting equal to zero in (\ref{eq:sch_e}):
\begin{equation}\label{eq:sch_ph}
    2 f - r f' = 0.
\end{equation}

Particularly, for the vacuum Schwarzschild solution with $f=1 - 2M/r$, from (\ref{eq:sch_e}), (\ref{eq:sch_de}) and (\ref{eq:sch_ph}) we get well-known results
\begin{equation}
    \E_\pm^2/m^2 = \frac{(r-2 M)^2}{r (r-3 M)},
    \qquad
    r_{ISCO} = 6M,
    \qquad
    r_{PS} = 3M.
\end{equation}

Similarly, Reissner-Nordström solution with function $f=(r^2 - 2Mr + Q^2)/r^2$ leads to the modified energy expression for neutral particles:
\begin{equation}
    \E_\pm^2/m^2 = \frac{\left(r (r-2 M)+Q^2\right)^2}{r^2 \left(r (r-3 M)+2 Q^2\right)}.
\end{equation}
Here, $Q$ is the electric charge of the background black hole. The radius of ISCO orbit must satisfy the following condition
\begin{equation}\label{eq:rn_de}
    M r^2 (r - 6 M) + 9 M Q^2 r - 4 Q^4  = 0,
\end{equation}
while the radius of the photon surface is
\begin{equation}
    r_{PS} = \frac{1}{2}\left(3 M + \sqrt{9 M^2-8 Q^2}\right).
\end{equation}
In order to compare ISCO radius with results of Ref., \cite{Junior:2024cts} we solve Eq. (\ref{eq:rn_de}) for $Q=0.2$:
\begin{equation}
    r_{ISCO}/M \approx 5.93957.
\end{equation}
At this point we fully agreed with the results of Ref. \cite{Junior:2024cts}, however in subsequent examples we find significant differences in our calculations.

To confirm the correctness of the full expression (\ref{eq:energy_g}), let us compare it with the known result\cite{Pugliese:2013xfa}. To describe the charged massive particles surfaces in the Reissner-Nordström metric we take the vector potential in the form
\begin{equation} \label{eq:rn_potential}
  A_t = - Q/r, \quad F_{rt}=Q/r^2,
\end{equation}
Similarly to  Ref.,\cite{Kobialko:2022uzj} we find the energy of the electrically charged particle in Reissner-Nordström metric:
\begin{align} \label{eq:rn_energy}
    \frac{\E_\pm}{m} &= 
    \frac{qQ}{2 m r}+
   \\ &+ \frac{
            (qQ/m) \left( Q^2 - M r \right)
            \pm 2 \Delta \sqrt{ \Delta - M r + Q^2 + (qQ/2m)^2 }
        }{
            2 r \left(r^2-3 M r+2 Q^2\right)
        }.\nonumber
\end{align}
Despite the fact that (\ref{eq:rn_energy}) seems at first glance different from,\cite{Pugliese:2013xfa} it gives exactly the same result. For example for $M = 1$, $Q = 0.9$, $q = -0.1$, $r = 10$, $m = 1$ we find numerically $\E_+=0.950969$. This numerical expression coincides with $E_+$ in Ref.,\cite{Pugliese:2013xfa} Eq. (9) with the maximal machine accuracy.  

\subsection{Electrically charged dilatonic black holes}

The metrics for electrically charged dilatonic black holes  can be written in the form \cite{Julie:2017rpw,Heydari-Fard:2021pjc}
\begin{align}
    \alpha&=\lambda^{-1}=\left(1-\frac{r_+}{r}\right)\left(1-\frac{r_-}{r}\right)^{\frac{1-a^2}{1+a^2}},\\
    \beta&=r^2\left(1-\frac{r_-}{r}\right)^{\frac{2a^2}{1+a^2}},\qquad
    \gamma=\beta\sin^2\theta,
\end{align}
with the vector potential
\begin{align}
A_t= -\frac{Qe^{2a\varphi_\infty}}{r}, \quad F_{rt}=\frac{Qe^{2a\varphi_\infty}}{r^2}.
\end{align}
Putting the metric and the field in (\ref{eq:energy_g}), we get the general energy expression for a charged massive particle:
\begin{widetext}
\begin{align} \label{eq:EMD}
    \E_\pm =&\left(1- \frac{1}{2+ \frac{3a^2-1}{1+a^2} \frac{1}{r/r_--1} -\frac{1}{r/r_+-1}}\right) \frac{qQe^{2a\varphi_\infty}}{r}\pm
\nonumber\\& \pm m \left(
              \frac{ 2\left(1-\frac{r_+}{r}\right)\left(1-\frac{r_-}{r}\right)^{\frac{1-a^2}{1+a^2}} \cdot \left(1+\frac{a^2}{1+a^2} \frac{1}{r/r_--1}\right)}{ \left(2+ \frac{3a^2-1}{1+a^2} \frac{1}{r/r_--1} -\frac{1}{r/r_+-1}\right)}
            + \frac{ \frac{q^2Q^2e^{4a\varphi_\infty}}{m^2r^2} }{\left(2+ \frac{3a^2-1}{1+a^2} \frac{1}{r/r_--1} -\frac{1}{r/r_+-1}\right)^2} 
        \right)^{1/2}.
\end{align}
\end{widetext}
In particular, the photon surface can be found from the following quadratic equation (vanishing denominator in Eq. (\ref{eq:EMD})):
\begin{align}
2 (1 + a^2) r^2 - 3 (1 + a^2) r r_+ - (3 - a^2) r r_-+ 4 r_- r_+ = 0.
\end{align}
It is easy to see that this expression defines the same photon surface as the expression (25) in Ref. \cite{Heydari-Fard:2021pjc}. Thus, this result is consistent with the geodesic approach.

\subsection{Conformal gravity}

Let us consider the first example that was mistakenly taken in Ref. \cite{Junior:2024cts} as a signal of the inconsistency of the MPS formalism. Schwarzschild solution in conformal gravity Ref. \cite{Mannheim:2011ds} has the form of the usual vacuum Schwarzschild solution with an extra factor (see more details in Refs. \cite{Junior:2024cts,Bambi:2016wdn}): 
\begin{align}\label{eq:conformal_gravity}
 &\alpha = \Omega^2 f, \quad
 \beta =\Omega^2 r^2, \quad
 \gamma = \Omega^2 r^2 \sin^2\theta,\quad
 \lambda = \Omega^2 f^{-1},
\end{align}
where $\Omega^2 = 1 + {l^4}/{r^4}$ and $f = 1 - 2M/r$.
Substituting this expressions in (\ref{eq:energy_q_zero}), we easily obtain
\begin{align}  
  \frac{\E_\pm^2}{m^2} &=
    \frac{ \left(r-2 M\right)^2}{ r(r-3 M)}\cdot  \left(1-\frac{l^4}{r^4}\right). 
\end{align} 
This is the same expression as in the geodesic approach (see Ref., \cite{Junior:2024cts} Eq. (78a)). Setting the denominator to zero, gives us the photon sphere $r_{PS}=3 M$, which does not depend on parameter $l$ as expected. Condition $d\E/dr = 0$ which gives ISCO orbits can be rewritten in the form:
\begin{align}  
\left(4 r^2-21 M r+30 M^2\right)l^4+r^4 (r-6 M)M=0.
\end{align}  
This expression is exactly the same as in the geodesic approach of Ref., \cite{Junior:2024cts} Eq. (79). In the Schwarzschild limit $l=0$, ISCO orbit is $r=6M$, as expected. In order to compare the result with one obtained in Ref. ,\cite{Junior:2024cts} we set $l=0.9$ and evaluate the ISCO radius $r=5.97564 M$, which accurately reproduces the result obtained with  geodesic approach in Ref. \cite{Junior:2024cts}.

\subsection{Culetu model}
 
Culetu model represents general relativity coupled to nonlinear electrodynamics Refs. \cite{Culetu:2013fsa,Culetu:2014lca}. This model has a black hole solution considered in Ref. \cite{Simpson:2019mud}. The solution has a relatively simple form Eq. (93) in Ref. \cite{Junior:2024cts,Simpson:2019mud}.  However, as it was noticed in Refs., \cite{Novello:1999pg,Toshmatov:2021fgm} description of the particle motion within this model requires using the effective metric presented in Ref.,\cite{Junior:2024cts} Eq. (98) in the following way: 
\begin{subequations}
\begin{equation}  
    \alpha =\frac{e^{-\frac{Q^2}{M r}} \left(r e^{\frac{Q^2}{2 M r}}-2 M\right) \left(32 M^2 r^2-14 M Q^2 r+Q^4\right)}{32 M^2 r^3},
    \end{equation} 
    \begin{equation} 
    \beta =-\frac{\left(r e^{-\frac{Q^2}{2 M r}}\right) \left(Q^2-8 M r\right)}{8 M},
    \end{equation} 
    \begin{equation}
    \gamma =-\frac{\left(r e^{-\frac{Q^2}{2 M r}}\right) \left(Q^2-8 M r\right)}{8 M}\sin^2\theta,
    \end{equation} 
    \begin{equation}
    \lambda =-\frac{32 M^2 r^2-14 M Q^2 r+Q^4}{64M^3r - 32M^2r^2e^{\frac{Q^2}{2 M r}}},
\end{equation}  
\end{subequations}
with the only non-zero elements of the field tensor and vector potential equal to
\begin{subequations}
\begin{align}  
    &
    F_{rt} =\frac{Q e^{-\frac{Q^2}{2 M r}} \left(Q^2-8 M r\right)}{8 M r^3},
    \\ &
    A_t=\frac{e^{-\frac{Q^2}{2 M r}} \left(Q^2-6 M r\right)}{4 Q r},
\end{align}  
\end{subequations}
where $Q$ is the electric charge of the black hole. Also, we will use $q$ for the particle's electric charge. 

Putting the metric and the field in (\ref{eq:energy_g}), we get the  general energy expression for a charged massive particle:
\begin{widetext}
\begin{align} 
\E_{\pm}&= \frac{e^{-\frac{Q^2}{2 M r}} \left(r e^{\frac{Q^2}{2 M r}}-2 M\right) \left(\pm\sqrt{\frac{D}{Mr}}-q Q \left(Q^2-8 M r\right)^2\right)R_3}{8M r \left(R_1-r^2  e^{\frac{Q^2}{2 M r}} R_2\right)}-\frac{q e^{-\frac{Q^2}{2 M r}} \left(Q^2-6 M r\right)}{4Q r},
\end{align}
where
\begin{subequations}
\begin{align} 
D &= M q^2 Q^2 r \left(Q^2-8 M r\right)^4 + m^2 \left(-32 M^2 r^2-6 M Q^2 r+Q^4\right)\left(R_1-r^2e^{\frac{Q^2}{2 M r}} R_2 \right), \\
R_1&=1536 M^4 r^4-1280 M^3 Q^2 r^3+308 M^2 Q^4 r^2-30 M Q^6 r+Q^8,\\
R_2&=512 M^3 r^3-368 M^2 Q^2 r^2+60 M Q^4 r-3 Q^6,\\
R_3&=32 M^2 r^2-14 M Q^2 r+Q^4.
\end{align}
\end{subequations}
\end{widetext}
The photon sphere determined from the condition of vanishing denominator gives the following equation:
\begin{align} 
R_1=r^2  e^{\frac{Q^2}{2 M r}} R_2.
\end{align} 
In particular, for $Q=0.5$ we numerically find $r=2.842M$ which accurately reproduces the  geodesic result Ref. \cite{Junior:2024cts}. For an uncharged $q=0$ massive particle, the squared specific energy has the form   
\begin{align}  \label{eq:uncharged}
\frac{\E^2}{m^2}=\frac{e^{-\frac{Q^2}{M r}} \left(r e^{\frac{Q^2}{2 M r}}-2 M\right)^2\left( -32 M^2 r^2-6 M Q^2 r+Q^4\right)R^2_3}{64 M^3 r^3\left(R_1-r^2  e^{\frac{Q^2}{2 M r}} R_2\right)}.
\end{align} 
At first glance, it may seem to differ from the expressions Eqs. (100), (101) in Ref. \cite{Junior:2024cts} obtained with the geodesic approach, but in fact they are the same up to some algebraic manipulations inside the expression. In order to verify this in a demonstrative way, we perform a numerical evaluation of $\E$ for $M=1$, $Q= 0.5$ and $r=5$ given by Eq. (\ref{eq:uncharged}) and Ref. \cite{Junior:2024cts} (Eqs. (100) and (101)). Both expressions give exactly the same value $\E/m= 0.932093$, which contradicts the main statement of Ref. \cite{Junior:2024cts}. 

We have shown complete consistency between the geodesic approach with the geometric approach of massive particle surfaces in all reported cases. Using the completely geometric approach, we immediately obtained already simplified expressions, which are fruitful for the subsequent task of visualization and analytical research.

\section{Conclusion}

We have described in detail the use of a new tool -- massive particle surfaces -- in the case of static spacetime. These surfaces are timelike hypersurfaces containing particles that never leave them if they are initially pushed in a tangential direction with proper energy. As special cases, they contain photon spheres and their generalizations. This method allows one to construct observables arising from the theory of strong lensing and black hole shadows without integrating geodesic equations. Our goal was to demonstrate that this method leads to results that coincide with the conventional geodesic approach, but provides certain advantages and opens new perspectives in the understanding of Killing tensor symmetries, uniqueness theorems, and Penrose-type inequalities. The second fundamental form of hypersurfaces plays a decisive role. We have clarified the useful concept of partial umbilicality of the MPS and presented expressions for the conserved energy in terms of its geometric parameters (\ref{eq:energy}) in the static case. We have demonstrated that the previously reported  in Ref. \cite{Junior:2024cts} discrepancies  with the geodesic approach are most probably due to incorrect expression for mean curvature of MPS used there. We hope that the explanation given in this paper will be useful to all researchers in this field to explore the new geometric approach and its possible applications.

\begin{acknowledgments}
This work was supported by Russian Science Foundation under Contract No. 23-22-00424.
\end{acknowledgments}

\appendix
\section{Conventions}
\label{sec:kh}

To avoid possible confusion with signs, we provide some basic expressions. The Lagrangian of the geodesic system reads
\begin{align}\label{eq:action_m}
S=\frac{1}{2}\int \{\sigma^{-1} g_{\alpha\beta}\dot{\gamma}^\alpha \dot{\gamma}^\beta -m^2 \sigma+2q A_\alpha\dot{\gamma}^\alpha\} ds,
\end{align}
Variations of this Lagrangian with a choice of parameterization $\sigma=1$ give
\begin{align}
\dot{\gamma}^\lambda \LCM_{\lambda}\dot{\gamma}^\alpha=q F^\alpha{}_{\lambda}\dot{\gamma}^\lambda, \quad g_{\alpha\beta}\dot{\gamma}^\alpha \dot{\gamma}^\beta=-m^2. 
\end{align}
The total particle energy associated with the Killing vector $k^\alpha$ is
\begin{align}
\E=-k^\alpha \pi_\alpha=-k^\alpha(g_{\alpha\beta} \dot{\gamma}^\beta+q A_\alpha).
\end{align}
Note that for the Reissner-Nordström metric with potential convention (\ref{eq:rn_potential}), we obtain the expression for the potential energy
\begin{align}
\E_p=q Q/r,
\end{align}
which has a sign of the usual form, for example, such as (3) in Ref. \cite{Pugliese:2013xfa}. However, other conventions may lead to different choices of sign. For example, as in Ref.,\cite{Kobialko:2022uzj} which should be kept in mind when comparing results.

~

~

\bibliography{main}

\end{document}